\def\year{2024}\relax
\documentclass[letterpaper]{article} 

\usepackage{arabtex}
\usepackage{utf8}
\setcode{utf8}

\usepackage{aaai22}  
\usepackage{times}  
\usepackage{helvet}  
\usepackage{courier}  
\usepackage[hyphens]{url}  
\usepackage{graphicx} 
\usepackage{natbib}  
\usepackage{caption} 
\DeclareCaptionStyle{ruled}{labelfont=normalfont,labelsep=colon,strut=off} 
\usepackage{algorithm}
\usepackage{algorithmic}
\usepackage{booktabs}
\usepackage{newfloat}
\usepackage{listings}
\floatstyle{ruled}
\newfloat{listing}{tb}{lst}{}
\floatname{listing}{Listing}

\usepackage{subcaption}
\usepackage[hang,flushmargin]{footmisc}

\title{Cross-Language Evolution of Divergent Collective Memory Around the Arab Spring}
\author{
    H. Laurie Jones,
    Brian C. Keegan,
}
\affiliations{
    Laurie.Jones@colorado.edu, Brian.Keegan@colorado.edu
}

\begin{document}
\setlength{\parskip}{0pt}

\maketitle

\begin{abstract}
The Arab Spring was a historic set of protests beginning in 2011 that toppled governments and led to major conflicts. Collective memories of events like these can vary significantly across social contexts in response to political, cultural, and linguistic factors. While Wikipedia plays an important role in documenting both historic and current events, little attention has been given to how Wikipedia articles, created in the aftermath of major events, continue to evolve over years or decades. Using the archived content of Arab Spring-related topics across the Arabic and English Wikipedias between 2011 and 2024,
we define and evaluate multilingual measures of event salience, deliberation, contextualization, and consolidation of collective memory surrounding the Arab Spring. Our findings about the temporal evolution of the Wikipedia articles' content similarity across languages has implications for theorizing about online collective memory processes and evaluating linguistic models trained on these data.
\end{abstract}

\section{Background}

Collective memory is the communal act of sense-making that results in community based perspectives of the past~\cite{halbwachs_collective_1992,zubrzycki_comparative_2020}. It is difficult to study empirically across linguistic and cultural contexts due to diverse evidence and sense-making processes. Wikipedia has become a test case, a ``global memory place'', to understand and compare the emergence and evolution of collective memory processes~\cite{pentzold_fixing_2009, yasseri_chapter_2022}. As a test case, the Arab Spring is major demonstration of internet mediated collective action that spanned across linguistic barriers and has become a salient case study to understand collective memory on Wikipedia~\cite{ferron_collective_2011,ferron_wikirevolutions_2011,al-shehari_negotiating_2022}

There are no requirements that different Wikipedia language editions have identical or even similar content about a topic. Because the content of each edit made to Wikipedia articles are stored in revision histories that can be retrieved and analyzed, it is possible to retrieve and compare across historical versions of articles. Researchers can leverage this temporal variation in article content within and across languages to identify the emergence of consensus perspectives, conflicts, and other processes associated with the construction and contestation of collective memory processes~\cite{ford_writing_2022}. This paper examines the construction of collective memory on Arabic and English Wikipedia articles about the events related to the 2011 Arab Spring by focusing on the use of two types of links within articles. The first type of links are ``outlinks'' from one article to another (also known as ``blue links'') that are strong signals of similarity between articles based on editors' judgments of relevance and context.\footnote{\url{https://en.wikipedia.org/wiki/Wikipedia:Manual_of_Style/Linking}}
The second type of links are inter-lingual links (ILL) identifying similar topics in other language editions.\footnote{\url{https://en.wikipedia.org/wiki/Help:Interlanguage_links}} 
This paper compares these links between articles across languages to detect \textit{what} is introduced (or removed) and \textit{when} as proxies for collective memory processes.

This paper contributes to two related empirical literatures using Wikipedia data. The first contribution is the study of different Wikipedia language editions by comparing their structure~\cite{porter_visual_2020,hickman_understanding_2021,adar_information_2009,he_the_tower_of_babeljpg_2018, dandala_towards_2012}, content~\cite{hecht_tower_2010, roy_information_2022,massa_manypedia_2012}, and collaboration practices like deliberation~\cite{hale_multilinguals_2014}. This literature is often focused on macroscopic patterns involving the entire project and comparisons at a particular point in time instead of more situated cases or the evolution of content. The second contribution uses Wikipedia data to understand collective memory processes~\cite{twyman_black_2017,luyt_wikipedia_2016,ferron_beyond_2014}. This literature is often focused on the collaborations and content in the days and weeks immediately following major events rather than collective memory processes that may unfold over years or decades. To address each of these limitations, we focus on a sample of articles related to the events of the 2011 Arab Spring that unfolded over 13 years ago to characterize and compare the ``afterlives'' of content about major historical events.


This paper presents an ensemble of quantitative methods to develop a grounded understanding of four behaviors related to collective memory formation: salience, deliberation, contextualization, and consolidation.


 The Arab Spring is what revolutionized our understanding of contemporary collective action. Our more complex definition of collective memory analyzing the Arab Spring summary page in English and Arabic, will allow a more thorough understanding of the entire collective memory process that is presented on Wikipedia. 

To analyze the salience, deliberation, contextualization, and consolidation of collective memory around the Arab Spring we identified four questions to motivate our work: 
\begin{enumerate}
    \item Is the concept salient through the time period in question, pointing to a continuing collective memory process? 
    
Through analysis of size and number of outlinks, we find that the Arab Spring is a salient across languages from 2011 until early 2024 (present). 

    \item Are results of varying deliberation processes across languages, creating divergent perspectives on the Wikipedia articles?

By clustering outlinks by their temporal inclusion, we find there are varying deliberation processes that are highlighted through previously latent inclusion themes we label, 'Stable', 'Debated', and 'Forgotten'.

    \item How similar are the outlinks contextualizing the event across languages over time?

Leveraging the ILLs of outlinks, we see that not only are the outlinks available across languages is different but that the majority of outlinks that are included post-phenomenon are isolated within each linguistic version of this article. 

    \item When do articles about related concepts link to the Arab Spring article, consolidating it within a broader context? 

Analysis of Ego-networks shows that in English articles about countries that are involved with the Arab Spring include reference to the Arab Spring but they do not in Arabic. Both languages debate the reflective reference to the phenomenon page when broken down into individual events within the Arab Spring. 

\end{enumerate}

\begin{figure*}[t]
\centering
\begin{subfigure}[t]{0.49\textwidth}
    \includegraphics[width=\columnwidth]{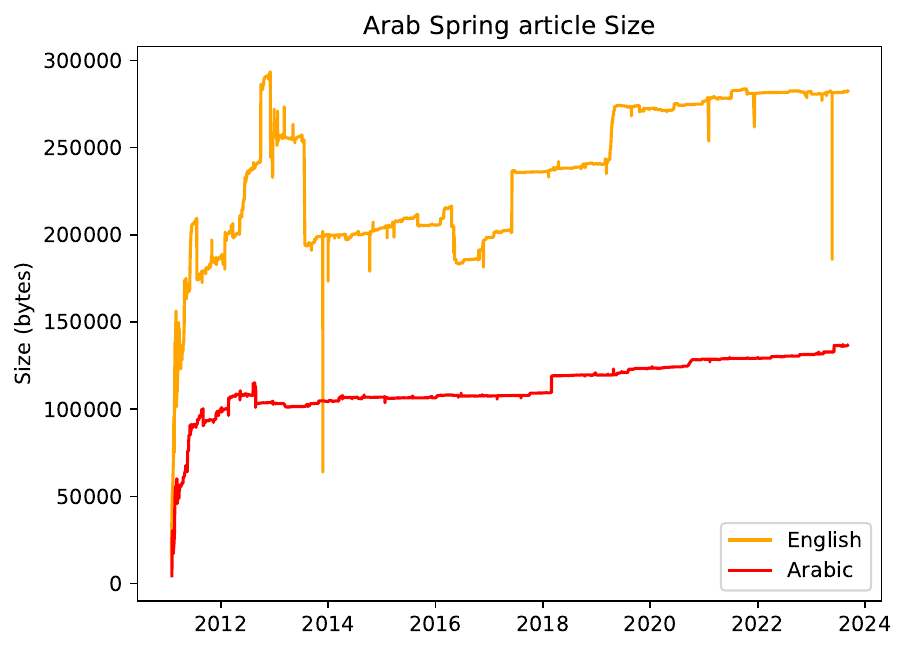} 
    \caption{Article size (bytes).}
    \label{fig:size}
\end{subfigure} \hfill
\begin{subfigure}[t]{0.49\textwidth}
    \includegraphics[width=\columnwidth]{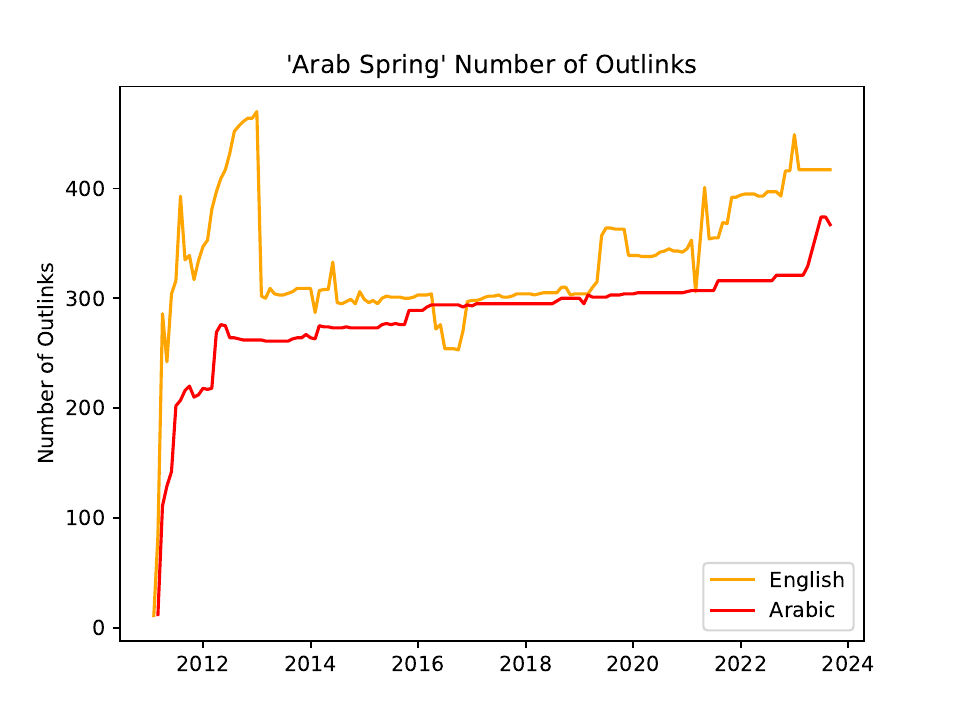}
    \caption{Outlinks.}
    \label{fig:outlinks}
\end{subfigure}
\caption{Changes in the size (left) and number of outlinks (right) for the ``Arab Spring'' article in English and Arabic used to operationalize salience of the article through time.}
\end{figure*}
\section{Longitudinal Measures of Event Salience}

Adding temporal complexity to collective memory, requires the topic to be salient through said time period. Within the context of Wikipedia, if a article has been neglected with little to no additions or changes, we assume the collective memory surrounding the topic to be settled and solidified within that linguistic community. To understand if the Arab Spring article is still salient, we looked at the size and number of outlinks within the article over time. Both of these attributes are simple measurements to identify continual efforts by editors to update and upkeep the article within contemporary contexts, possibly altering the way in which the Arab Spring is being remembered. For the sake of privacy and safety we will not disclose the names of the editors that make notable changes in our analysis. 

\subsection{Size}

According to~\cite{roy_topic-aligned_2020} about 65\% of articles are longer in English than in Arabic meaning the overall discrepancy in size between the two articles in not unexpected. Both articles continue to change over time, point to an active editor group, constantly re-contextualizing the article. However, upon further analysis we realize that the size of the article is also heavily influenced by the editing of external references\footnote{External references refer to links to websites outside of Wikipedia, predominantly seen in sources}.

In the English Arab Spring summary article the size has large variations at certain points in time. The large drop in 2012, was due to the transition of this article into a summary page as the outlinks were deleted from here and moved to the article about ‘concurrent incidents’. The next drop in 2016, as mentioned was due to an edit of references, but was influenced also by a restructuring of the ‘Social Media and the Arab Spring’ sub-section of the article. The size then started to rise a few months later from its local minimum, by editors adding to 'Aftermath' and 'Social Media' subsections of the article. Then in the beginning of 2017 a bot ‘fixed 198 sources’ largely increasing the size. The last large change in size was due to an individual editor, who in 2019 came in and restructured most of the page, adding or altering 5 sub sections of the article. These large changes in size through time points to continued salience within the editor community as they maintain external links as well as update subsections particularly about social media and what happened after the Arab Spring. 

With only two main increases in size the Arabic Arab Spring summary article, a picture of an article that has very little negotiation or updating after 2013  is presented. In 2013, an editor deleted interwiki links\footnote{Interwiki links can be a link to another project, to another language and both, to another project in another language.} away the article similar to external links having a relatively large change in size. In 2018, a bot repaired external references that linked to web pages that no longer existed. Although there isn't large variation in size like the English Arab Spring article, its consistent increase in size is due to editors continually making changes with edits even in 2023 such as,

\RL{"الجملة فيها مقصد مستفز لبعض السوريين المؤيدين للحكومة السورية"}
(This sentence has a provocative meaning for some Syrian supporters of the Syrian government) resulting in a decrease of 19 bytes and \RL{"معلومات دون مصادر"} (Information without sources, removing tags) resulting in an increase of 63 bytes, constantly re-adjusting small things but that is related to the collective memory process. 
\vspace{-.45cm} 
\subsection{Number of Outlinks}
\vspace{-.45cm} 
Revealing additional examples of salience, number of outlinks captures other types of engagement that is related to the collective memory processes such as the linking of concepts that were previously un-linked. This provides additional information and context that is not available just looking at size.
In the English version of the Arab Spring Summary article, outside of the largest drop of outlinks in 2012, due to the restructuring resulting in the English Arab Spring summary page, single editors coming in and making large updates to the article are what influence the number of outlinks changing. For example, in 2016 two editors came in and restructured  the 'Social Media and the Arab Spring', 'International Pressures' and 'Aftermath' subsections of the article leading to the inverted plateau. The large increase in 2019 was the same single editor revision that caused an increase in size by altering 6 sub sections of the article that included, 'Results', 'Aftermath', 'Social Justice', and 'Counter-revolution and Civil Wars'. In 2021, someone tried to redo the summary box and although it was immediately undone, and in 2022 an editor added 23 links to the Outcomes section but only caused the article to increase by 3 bytes in size and was shortly undone not long after

In the Arabic version of the article, the largest increase in number of outlinks was in February of 2012, someone did comprehensive overhaul of the introduction. There are two relatively small increases in October of 2015 where someone added more information about the events in Jordan and July of 2021 when a bot and a person both added information about Morocco. The last large increase was in March of 2023 where an individual editor make a large contribution over the whole month that he primarily tagged, \RL{ تحرير مرئي } or 'Visual Editing'. 
The continual changes in size across languages and through time points to continued salience within both editor communities as they maintain external links and update subsections, particularly about social media and what happened after the Arab Spring in English. Although through smaller changes in Arabic, the continual engagement of editors through time confirms our question 1: the Arab Spring has continued to be a salient topic within both the English and Arabic Wikipedia communities. 



\section{Deliberation Through Temporally Clustered Vectors}
\begin{figure*}[t]
    \centering
    \includegraphics[width=2\columnwidth]{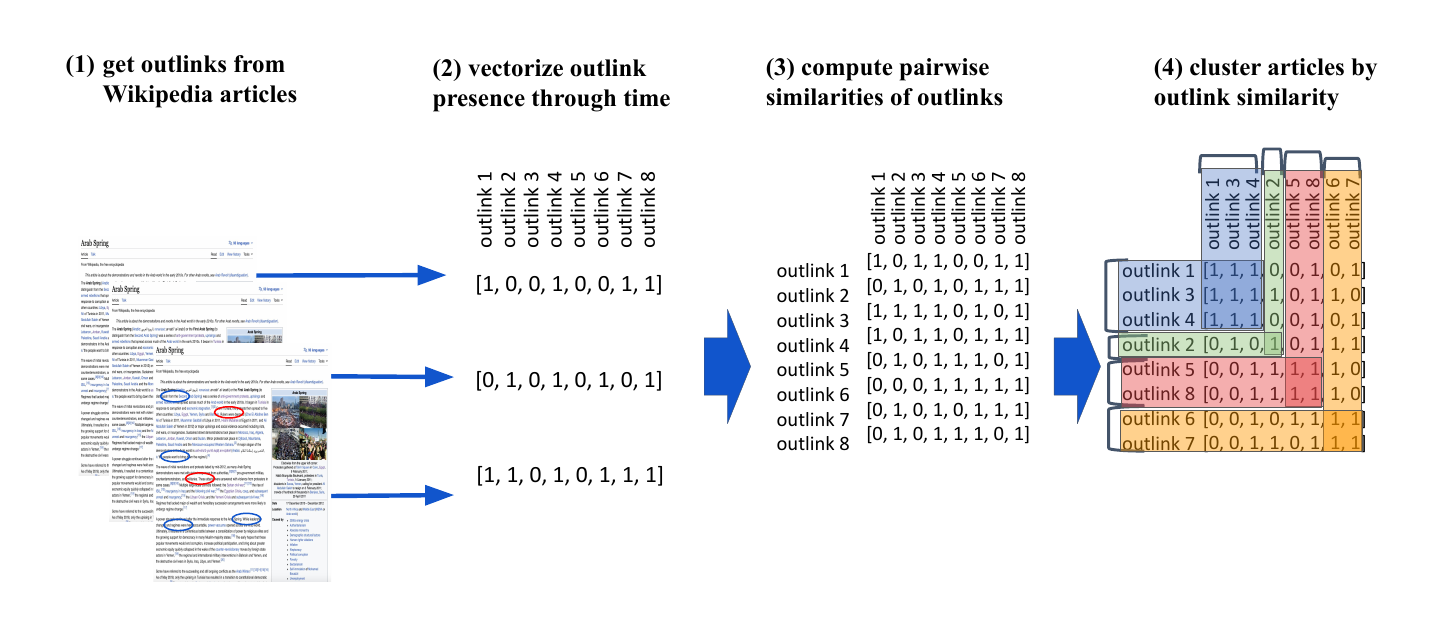}
    \caption{This is a figure breaking down the outlink temporal inclusion vectorization and clustering process. Beginning by identifying outlinks on each revision of the Wikipedia article and then a one hot encoding of the outlink inclusion in the English or Arabic Arab Spring articles, this results in a clustered heatmap, revealing groups of outlink that are later identified with stable, debated, and forgotten patterns}
    \label{fig:heatmapsummary}
\end{figure*}
Focusing on deliberation through editor perspectives and discussions has been the previous standard for analyzing collective memory ~\cite{ferron_collective_2011, keegan_dynamics_2019, hickman_understanding_2021, hale_multilinguals_2014}. Although this is an intuitive look at the the sense-making process, the production resulting from these discussions is what broader members of the linguistic community will interact with. That is why we operationalize deliberation through temporal outlink inclusion and exclusion. Outlinks are built into the Wikipedia infrastructure as a way to help readers learn more about a topic that an editor deems relevant to understanding the topic at hand. Their inclusion in the article can be the result of the previously studied editor dynamics but their inclusion and exclusion over time also lends insight into changing perspectives because information on Wikipedia articles are constantly being deliberated as new people get exposed to the content, as we saw through our analysis of article salience. 

To analyze these outlinks through temporal inclusion we implement one hot encoding and vectorize outlink inclusion for each revision version of the Arab Spring article. We clustered these concepts by their pairwise similarity scores to create a heatmap that positions words with similar scores closer together creating a more digestible heatmap. This resulting heatmap revealing the latent inclusion patterns of broader themes and topics.

\subsection{English}
There are the clusters of nodes going along the diagonal of the clustered heat map. These clusters are sub collections of outlinks that have similar temporal inclusion. These have then been broken down and analyzed by their subcollection, gaining a emergent label of 'Stable' if the inclusion pattern consistently increases in number major deletions, 'Disputed' if there is a consistent pattern of inclusion and exclusion or 'Forgotten' if those outlinks are primarily deleted and never added back again.

\begin{figure*}[t]
    \centering
    \begin{subfigure}[t]{0.49\textwidth}
        \includegraphics[width=\columnwidth]{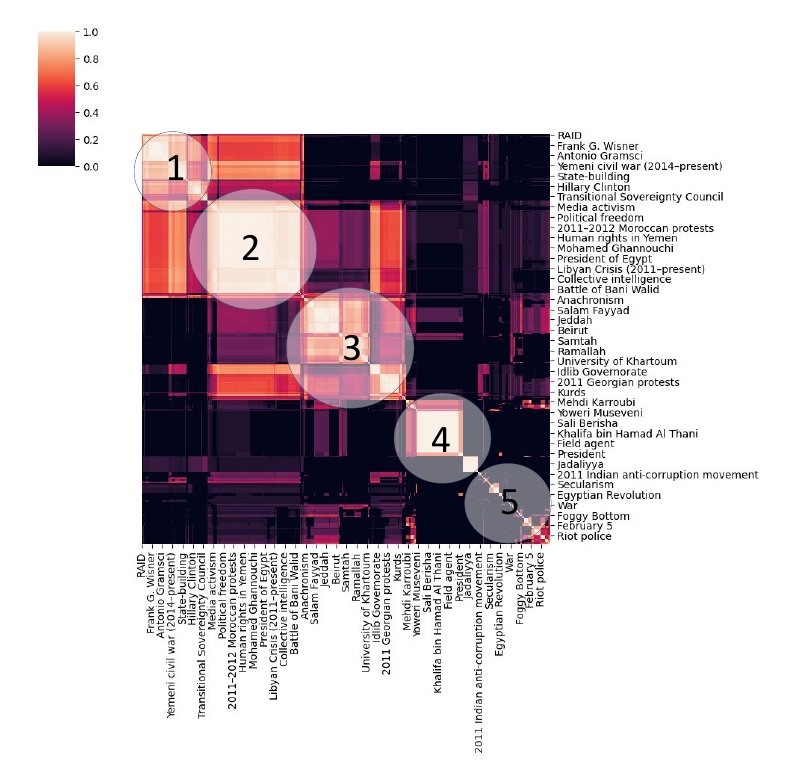}
        \caption{This is a pairwise similarity clustered heatmap of outlink temporal inclusion vectors for revision editions of the English Arab Spring article.}
        \label{fig:en_heatmap}
    \end{subfigure}\hfill
    \begin{subfigure}[t]{0.49\textwidth}
        \includegraphics[width=\columnwidth]{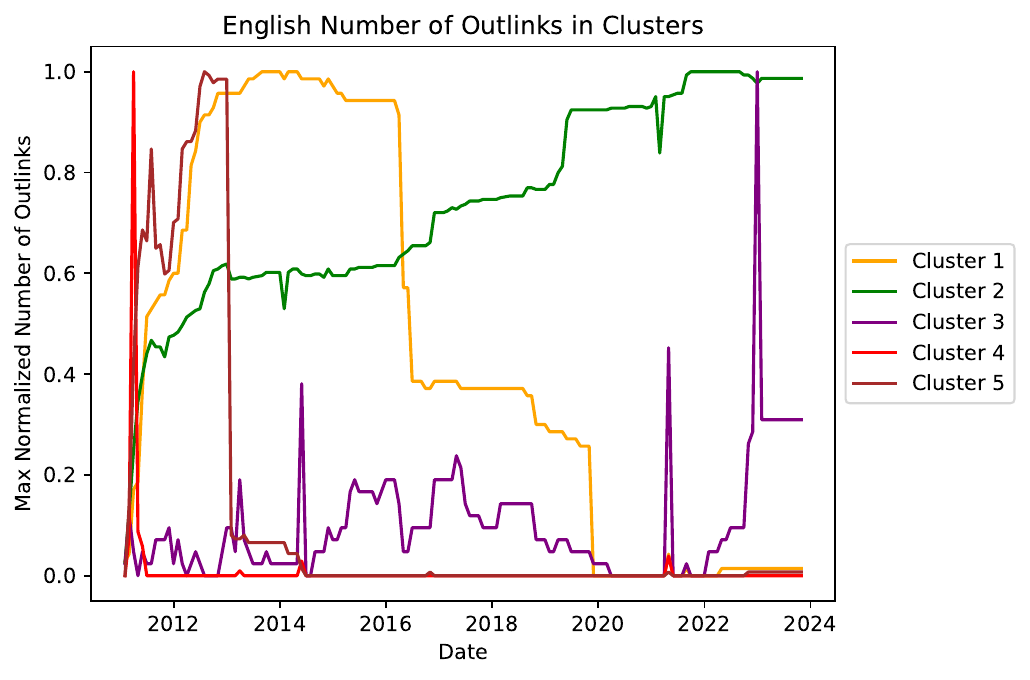}
        \caption{These are the number of outlinks included through time within each clusters of the English Arab Spring heatmap, the x-axis is normalized by maximum values}
        \label{fig:en_heatmap_clusters}
    \end{subfigure}
    \caption{English outlink similarities. }
\end{figure*}

\subsubsection{Cluster 1}
This cluster has identified what we label to be 'Forgotten' concepts because by 2020, all of the outlinks in this cluster were removed an are never added again. In 2016, outlinks such as “2009–10 Iranian election protests", “Bahraini uprising (2011–present)|Bahraini”, and "2011–12 Moroccan protests”, links to articles about smaller events were completely forgotten. An editor came in and deleted the sections on international reactions and external effect but added the section disparate outcomes, deleting previous connections to international reactions and altering the perspective of subsequent events. 

The second dip of this clustered outlink inclusion was the end of 2019 and mostly lost links about about Syria such as the "Syrian Army", and "Aleppo Governorate". This was due to an editor contemporarily re-contextualizing the article. This was done by altering sections about the aftermath, Yemen, Libya, and deleting a large description about Syria suggesting a move to the Syrian Civil War Wikipedia page. Now that information is no longer available to people who are looking for summaries of the events on the Arab Spring. 

\subsubsection{Cluster 2}
We give cluster 2 the label 'Stable' as these articles rarely have any deletions and make up a majority of the current version of the article. The number of articles within this subgroup consistently increases over time but what is added varies. This group includes foundational articles such as "2011 Egyptian revolution" and "2011 Syrian revolution" but as time goes on, more articles such as "Foreign policy", "Media Activism", and "Barack Obama" are added. Within the deliberation process, it is not-disputed that as time goes on, additional information about the west and outlinks we identify as related to 'Political Philosophy' concepts are increased.  

\subsubsection{Cluster 3}
Cluster 3 is a 'Debated' cluster. These topics are constantly being added and taken away. In 2014, outlinks such as "Civil Disobedience", "Defection", "Protest Camp" and "Rebellion", were added to the ‘Causes’ section of the Arab Spring article and then immediately undone. In 2021 topics such as the "2000s Energy Crisis", "Economic freedom", and "Silent Protest" were added but because of a large deletion that the editor made, it was also immediately undone. In 2022 topics such as the "2019 Gaza Economic Protests", "BBC" and "CNN" were added again by an anonymous person that immediately gets undone moments later. The large spike in 2023 sees many of the same outlinks in 2022, "BBC", "CNN", and once again these get deleted. However, some outlinsk such as "Insurgency" and "Khartoum massacre", an event that happened in 2019 stays. Not identified in the spikes are clusters of outlinks that are added for a few months, deleted and then added again. This includes links to pages such as the "Western Sahara" and content about India such as "2011 Indian Anti-Corruption Movement" "Parliament of India"  These are represented in the plateau shapes from 2015 to 2020.  

In all three of these cases there was an anonymous edit that added outlinks that was then undone within hours if not minutes with no discussion as to why.
The large anonymous edits, typically from mobile devices is often labeled as 'vandalism'.\footnote{\url{https://en.wikipedia.org/wiki/Vandalism_on_Wikipedia}} 


\subsubsection{Cluster 4 and 5}
We analyze Cluster 4 and Cluster 5 together. They both are sub-groups of 'Forgotten' concepts and both are a part of the active evolution of the Arab Spring phenomenon as everything in both clusters are removed by 2014. This is the outlink evolution that would be captured in previous literature that limits the temporal scope to the time of phenomenon activity. 

The large spike in Cluster 4 happened in the month of March in 2011. These were all included that month and then by the end of March 2011 they were deleted. Previously included were topics about other countries outside of the Middle East are forgotten such as "Chinese Communist Party", "Republic of Korea Armed Forces", "Socialist Party of Albania", "Uganda", "Zimbabqe" and "South Sudan".  This is a very early time in the Arab Spring phenomenon. This cluster reveals discussions and decisions about what this page and phenomenon is supposed to be about and the resulting excluded topics. Many of the references to other countries were about events that were happening simultaneously. For example, in February and March there is mention of leaflets about the Arab Spring dropped in North Korea by the South Korean military. This was subsequently moved to a different article titled 'Impacts of the Arab Spring'.  



Cluster 5 has a small decrease at the end of 2012 and completely minimized in 2014. This includes outlinks such as "Cabinet of Yemen" and "2011 Lebanese protests". Where Cluster 4 referrs to concepts outside of the Middle East, this cluster points to 'Forgotten' concepts within the Middle East, such as around Palestine ("Palestinian Intifada" "Gaza Strip"), the Gulf ("Royal Oman Police", "United Arab Emirates") and other concepts relevant to the region at the time such as "Media Blackout". This deletion was primarily due to an editor "splitting the Arab Spring into concurrent incidents" where they deleted the concurrent incidents section of this article and put it into a different Wikipedia page. This is an interesting choice because it deleted specifics about countries such as 'Algeria' but some details were still left in the overall page, included in descriptions of initial protests. This deletion further specifies what countries are considered 'officially' within the Arab Spring.

\subsection{Arabic}

\begin{figure*}[t]
    \centering
    \begin{subfigure}[t]{0.49\textwidth}
        \includegraphics[width=\columnwidth]{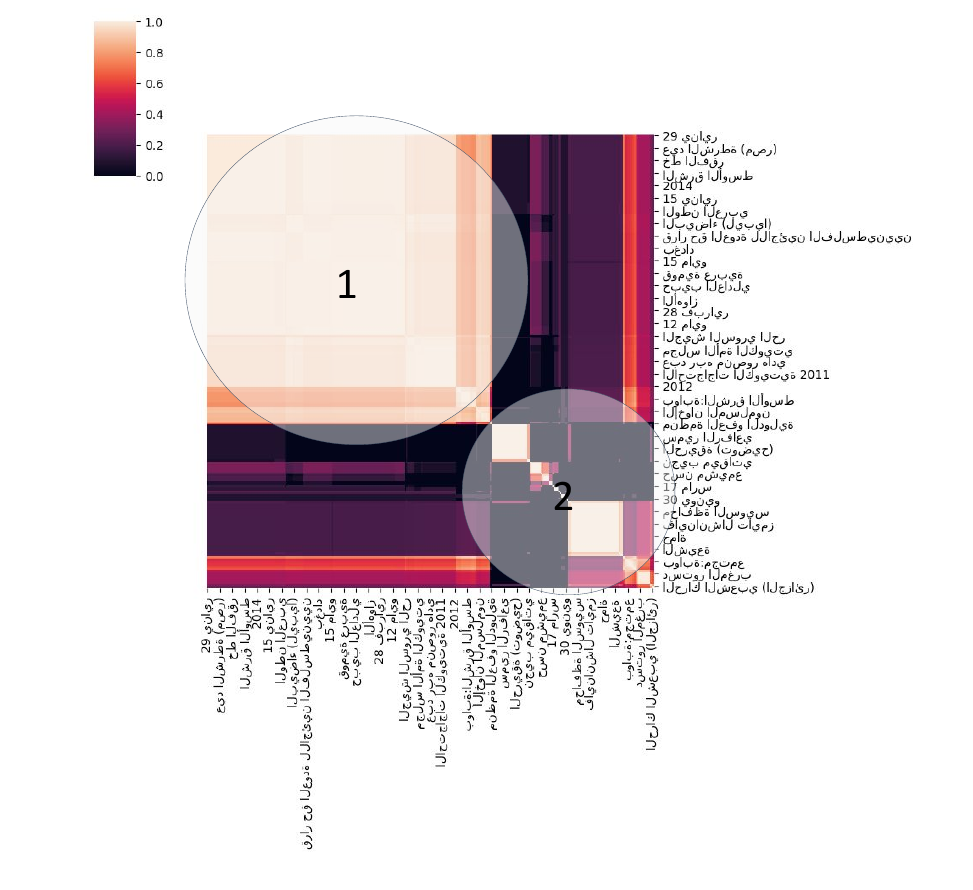}
    \caption{This is a pairwise similarity clustered heatmap of outlink temporal inclusion vectors for revision editions of the Arabic Arab Spring article.}
    \label{fig:ar_heatmap}
    \end{subfigure}\hfill 
    \begin{subfigure}[t]{0.49\textwidth}
        \includegraphics[width=\columnwidth]{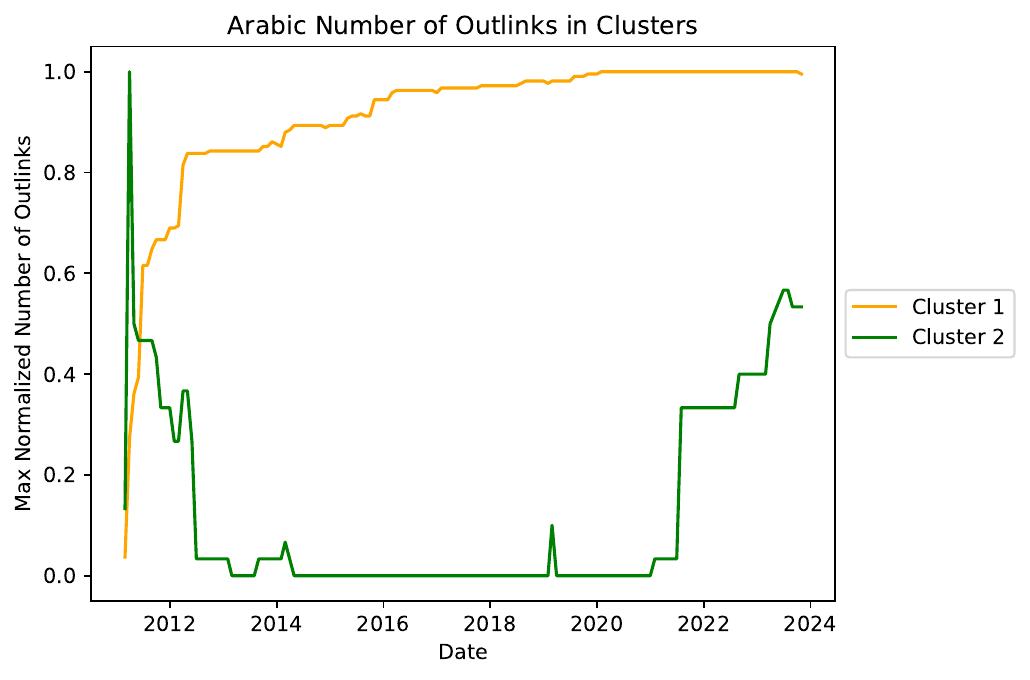}
    \caption{These are the number of outlinks included through time within each clusters of the Arabic Arab Spring heatmap}
    \label{fig:ar_heatmap_clusters}
    \end{subfigure}
    \caption{Arabic outlink similarities.}
    \label{}
\end{figure*}
    
\subsubsection{Cluster 1}
Similar to Cluster 2 in the English heatmap, these outlinks are considered "Stable". They make up a majority of the contemporary version of the article and also consistently increase in time. In addition to having the foundational links of the phenomenon, the Arabic language version also includes a large infrastructure of dates, for example, although \RL{يناير}14 (January 14) links to an article that describes everything that happened that day, it is also a major day of protests in Tunis, Lebanon, and Jordan. Outside of these links, as this article evolves this subsection begins to include concepts about contemporary events and conflicts. For example, in 2014, \RL{لجيش السوري الحر} (The Free Syrian Army) is added, in 2018, \RL{احتجاجات جنوب العراق 2018} (The Southern Iraqi Protests 2018) and in 2021, \RL{أحزاب سياسية في المغرب} (Political Parties in Morocco) are added. It is agreed upon by readers and editors to continue to update and contextualize this article within other events that happen after the initial phenomenon. 

\vspace{-3.5cm}
\subsubsection{Cluster 2}
Cluster 2 proposes primarily two distinct inclusion times, before 2014, and after 2021. There is actually no overlap between these two inclusion groups and those we label this cluster as both 'Forgotten' and 'Debated'. In the beginning, outlinks are all taken out by 2014, similar to Cluster 5 in the English article. In this Arabic version, concepts such as \RL{أوغندا}(Uganda), \RL{إثيوبيا} (Ethiopia), \RL{الشباب (تنظيم صومالي)} (Al Shabaab (Somali organization)), and \RL{حسن مشيمع} (Hasan Mushaima), an opposition leader in Bahrain. These are the 'Forgotten' concepts, outlinks that never get referenced again. These are mostly all taken out in a large edit by an editor that pointed out that the Arabic version is the only version that includes the protests in Somolia as a part of the Arab Spring and then deletes the related section and content.

Outlinks that are only at the end of this cluster are \RL{أحزاب سياسية في المغرب} (Political Parties in Morocco), and \RL{دستور المغرب} (Moroccan Constitution). Similar to Clusters 1, 3, and 5, in the English version of this article, this subgroups this appears to be a deliberation over what broader regional events should be included in the Arab Spring summary page. However, because they are not included until recently and all remain included in the current version, these are labeled as 'Debated'. They are added in this section by an editor with a small description in June of 2021. Without a description this editor re-writes the descriptions of the events in Morocco, excluding its ties to other protests and focusing more on internal events. 

Understanding outlinks to reflect distinct deliberation processes, we identify a 'Stable' inclusion of western concepts in the English version and contemporary conflicts in the Middle East in Arabic. 'Forgotten' and forgotten concepts related to simultaneous events in the Middle East, Asia, and Northern Africa in both languages, similar to the English version pointed to a definition of the Arab Spring is and the events that it encompasses. 'Debated' concepts within the editor and non-editor community in English that evolve as different editors get involved and contemporarily re-contextualize the phenomenon.

\section{Contextualization While Acknowledging Linguistic Access}

Focusing on collective memory as a sense-making process, contextualization reveals how editors position one topic when compared to others. Outlinks are clear approximations for contextualization by definition, but we must first understand the outlinks that are available to be used and how they are used. We leverage the inter-lingual links to identify resource discrepancy and silohed information. The Wikipedia API does not have temporal data for ILLs so we will have to classify their inclusion upon the current languages that outlinks are available in. Although not ideal, this allows us to address temporal resource variation within the broader Wikipedia linguistic ecosystem. 

\subsection{ILLs of Contemporary Outlink Contextualization}
To understand the temporal contextualization, we will first look at the current outlink statistics of the 'Arab Spring' article to put in perspective temporal characteristics.
\begin{table}[t]
    \centering
    \footnotesize
    \begin{tabular}{lcccc}
    \toprule 
        & \multicolumn{2}{c}{\textbf{English}} & \multicolumn{2}{c}{\textbf{Arabic}}  \\
        & \textit{Count} & \textit{Fraction} & \textit{Count} & \textit{Fraction} \\ \midrule
        \textbf{No ILL} &  18 & 5.5\%  & 21 & 7.8\% \\
        \textbf{ILL, no outlink} & 230 & 70.3\% & 169 & 62.8\% \\
        \textbf{ILL, outlinked} & 79 & 24.2\% & 79 & 29.4\% \\ \midrule
        \textbf{Total} & 327 & -- & 269 & -- \\ \bottomrule
    \end{tabular}
    \caption{The number of outlinks by inter-language link (ILL) status for the English and Arabic articles about the Arab Spring.}
    \label{tab:ill_freq}
\end{table}


Table~\ref{tab:ill_freq} shows that within both articles only around 25\% and 30\% of the 'Arab Spring' outlinks reference concepts that are shared between language versions of the article. A majority of these are links to either major conflicts or people related to the phenomenon such as "2011 Egyptian revolution", "Bashar al-Assad", "Muammar Gaddafi", and "Yemeni Revolution". 

A majority of the 18 English articles with no Arabic equivalent are American individuals and concepts such as "Zeynep Tufekci", an American professor studying social media of Arab Spring. 
 
Outlinks that are only in Arabic, fall into two main categories: 'structure-based' and 'content-based'. 'Structure-based' outlinks are links such as, \RL{بوابة:الشرق الأوسط} or "Portal: Middle East"\footnote{\url{https://en.wikipedia.org/wiki/Wikipedia:Contents/Portals}}
, and \RL{دستور تونس (توضيح)} or "Constitution of Tunisia (Disambiguation)"\footnote{\url{https://en.wikipedia.org/wiki/Wikipedia:Disambiguation}}
are articles that are made because of the structure of Wikipedia where as 'content-based' articles, in this subgroup are articles like \RL{المبادرة الخليجية} (Gulf Initiative), and \RL{محاكمة حسني مبارك} (Trial of Hosni Mubarak). These were made because people wanted to have articles about these topics.

A majority of the articles that have articles in both languages but only referenced in the English version, continue the pattern of  western concepts such as "Condoleezza Rice" as well as concepts about political philosophy such as "Authoritarianism" and "Economic stagnation". There are also additional outlinks to information across countries involved in the phenomenon. Such as, "Post-coup unrest in Egypt (2013–2014)", "Yemeni Crisis (2011–present)", and "2012–2013 Egyptian protests". 

There is a large infrastructure on Arabic Wikipedia of dates, and that is seen in the links that cross both languages but are only referenced in Arabic. For example, \RL{يناير 14} (January 14) links to an article that describes everything that happened that day; it is also a major day of protests in Tunisia, Lebanon, and Jordan. Other themes present in this section are topics about the Israel-Palestine conflict such as \RL{إسرائيل} (Israel) and \RL{حرب 1948} (1948 War) which English equivalent article is "1948 Arab–Israeli War" and outlinks themed around Government structures like \RL{مجلس النواب البحريني} which English article is, "Council of Representatives (Bahrain)". 

As seen in English language Cluster 2 of Deliberation, the 'Stable' outlinks are what culminate within the current version of the article. Both language articles shared foundational texts. There is a strong inclusion of western concepts and concepts about political philosophy isolated within the English version as well as available across languages but only referenced in English. Structural elements within the Arabic language edition of Wikipedia motivate the inclusion of certain outlinks but there are also many links that focus on Middle Eastern phenomena.

\subsection{Temporal Outlink Contextualization}
Building upon the contextualization of the contemporary version of the article, analyzing the ILL based outlink classifications through time reveals an evolving contextualization of the 'Arab Spring' article within the broader Wikipedia language ecosystem.  

\begin{figure*}[t]
    \centering
    \begin{subfigure}{.49\textwidth}
        \includegraphics[width=\columnwidth]{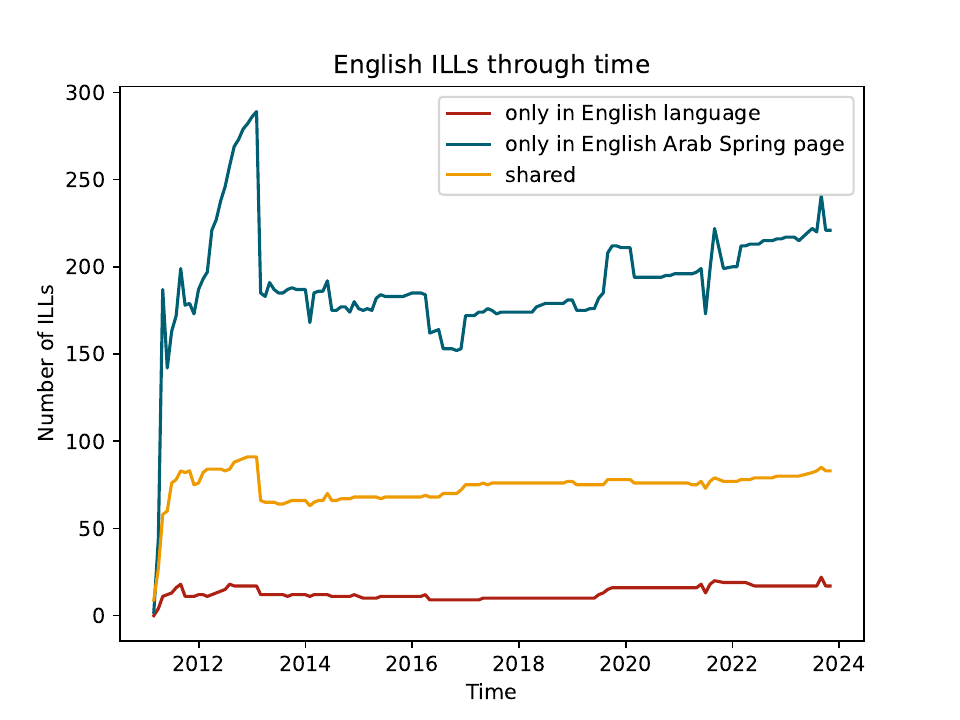}
        \caption{English inter-language links.}
        \label{fig:outlinks_en_to_ar}
    \end{subfigure}\hfill
    \begin{subfigure}{.49\textwidth}
        \includegraphics[width=\columnwidth]{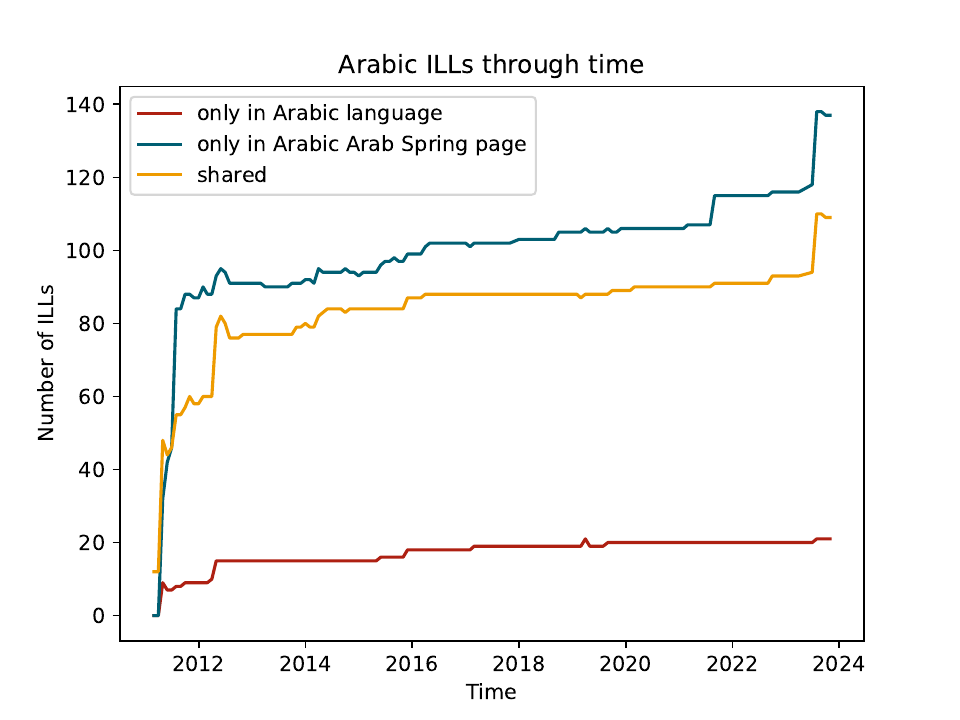}
        \caption{Arabic inter-language links.}
        \label{fig:outlinks_ar_to_en}
    \end{subfigure}
    \caption{Outlinks as they evolve through time in English(left) and Arabic(right) categorized by whether their current outlink has an ILL in the other language version of the article, and whether it is included in that other language version.This is an operationalization of the access and application of shared contextualization practices}
\end{figure*}

In figure 6(a), the drop in outlinks associated with the switching of the page to a summary page affects primarily the outlinks that have an ILL in both languages rather than the ones that are isolated in English. Most of the large changes in number of outlinks (figure 2), primarily stems from outlinks that cross languages but were only chosen to be included in the English version of this article. Since a majority of these edits revolved around the subsection of "Social Media" and sections about after the phenomenon, the English Arab Spring page contextualizes what happened after, primarily through an isolated perspective, referencing articles that are available in both languages but only referenced in this language edition. 

In figure 6(b) the Arabic Wikipedia page also reflects figure 2. The main change in outlinks from 2012 due to the change of the introduction, which makes sense because a majority of the shared articles are about foundational topics. However, the increase 2021 where a bot and a person made edits about Morocco appears to be isolated within concepts that are not referenced in the English article, providing an isolated linguistic perspective. Similarly the editor in March of 2023 who did alot of edits to various different parts did so by adding links that were already in the English page as well as add new links that would only be in the Arabic page. The number of shared outlinks between pages, is almost 50\% of the outlinks in this article at a given time is in part due to the size, although it is so few articles since the Arabic article is significantly smaller, foundational outlinks take up more of the article proportionally. This produces a contextualization process that is not does not vary far off of the foundational concepts. 



\section{Consolidation through Reference Dissemination}

How a community situates a topic in the broader memory system is reflected in how it is referenced when discussing other topics on Wikipedia, influencing how other memories are formed and contextualized. The consolidation of the Arab Spring as a fixture of this system would be its inclusion within the articles about the events and countries involved in the broader phenomenon. We identified eleven events\footnote{'2011 Omani protests', '2011–2012 Moroccan protests', '2011–2012 Jordanian protests', '2011 Iraqi protests', '2010–2012 Algerian protests', '2011 Bahraini uprising', 'Syrian revolution', 'Yemeni Revolution', 'Libyan civil war (2011)', '2011 Egyptian revolution', 'Tunisian Revolution'} that both the English and Arabic versions of the summary article deemed ‘important’ and the countries in which those events take place\footnote{These events and countries were cross referenced between languages and the only discrepancy is that in the English article, the event associated with Syria was the “Syrian Revolution” article which is not referenced in the Arabic article at all. Instead in the Arabic article, “The Syrian Civil War” page is referenced. Since that article is also referenced in the English article, that is the event we investigated.}.
We investigated this consolidation by generating country and event based ego networks\footnote{A type of network from social network analysis that maps connections from a particular perspective}
for monthly page revisions of the articles and their outlinks and measuring inclusion within those networks. The resulting figures do not present whether the 'Arab Spring' article is mentioned but whether editors consider the 'Arab Spring; summary article a necessary and relevant point of information to contextualize the country or event. 

\begin{figure*}[t]
    \centering
    \begin{subfigure}{.49\textwidth}
        \includegraphics[width=\columnwidth]{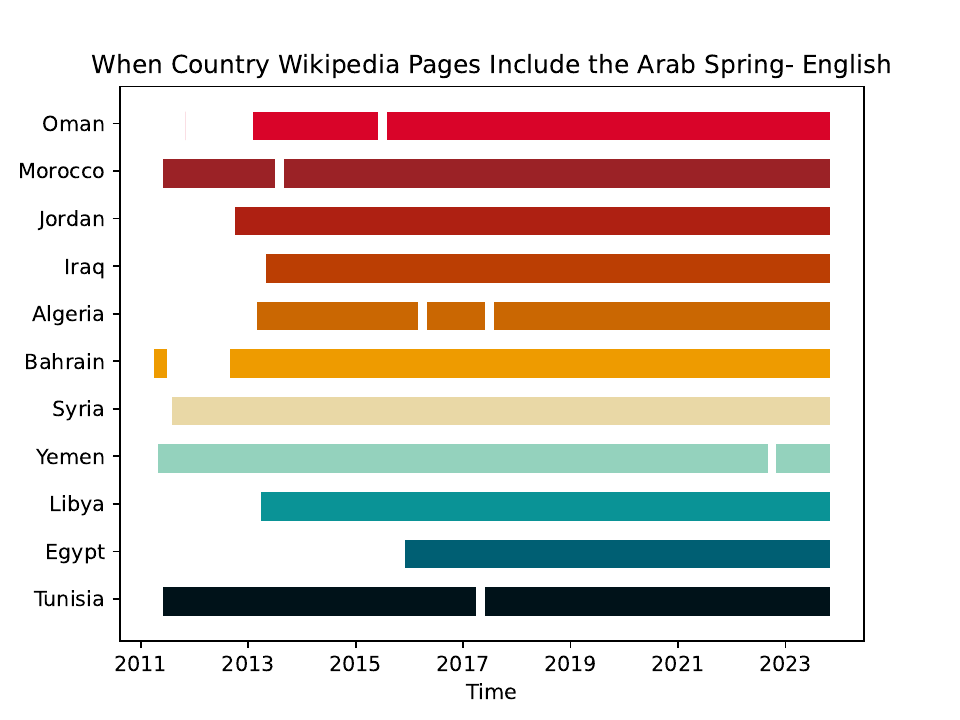}
        \caption{Country articles.}
    \end{subfigure} \hfill
    \begin{subfigure}{.49\textwidth}
        \includegraphics[width=\columnwidth]{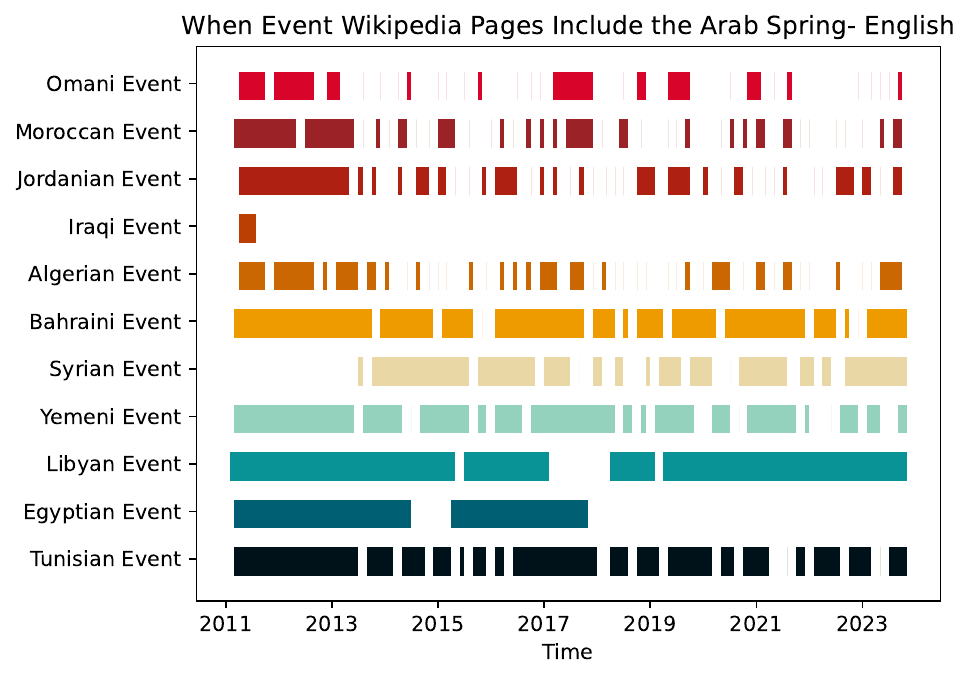}
        \caption{Event articles.}
    \end{subfigure}
    \caption{The first graph is a visual of the temporal inclusion of the 'Arab Spring' as an outlink in the English language country articles. The second graph is the temporal inclusion of the 'Arab Spring' in the English language event articles. This presents a perspective on the consolidation of the Arab Spring as a contextualization for the country and event Wikipedia articles.}
    \label{fig:temporal_inclusion_en}
\end{figure*}

Figure7(a) is a visual representation of the inclusion of the “Arab Spring” as an outlink within the English language country articles. The English language country articles appear to only have slight deliberation about the inclusion of the 'Arab Spring' in the articles. In 2014, all but one of the country's Wikipedia articles referenced the Arab Spring as an outlink in their article. This has a stark contrast to the English language event articles in Figure 7(b)that show debate about the inclusion of the 'Arab Spring' summary page.
The 'Arab Spring' appears to consolidate the most within the Libyan, Bahraini, and Yemeni context with both the country and event having the most inclusion through time. These events have quite different resolutions with events in Bahrain ending in 2018 and events that were triggered by the 'Arab Spring' creating diverting contemporary problems in Libya and Yemen.
The Iraqi, Egyptian, and Omani articles have the lowest inclusion in both their event and  country articles. This points to a delayed connection between to 'Arab Spring' and could point to a distinction between the Arab Spring and those events. Alternatively it could be so tied to the concept of the Arab Spring, an additional reference wouldn't be necessary.
There is a discrepancy between the 'Syria' article including the 'Arab Spring' consistently from the beginning and the Syrian event not beginning to include or negotiate its inclusion until after 2013. Within English Wikipedia, editors found the Arab Spring a defining characteristic of countries across the MENA\footnote{Middle East and North Africa} region and debate over phenomenon level referencing.

\begin{figure*}[t]
\centering
    \begin{subfigure}{.49\textwidth}
        \includegraphics[width=\columnwidth]{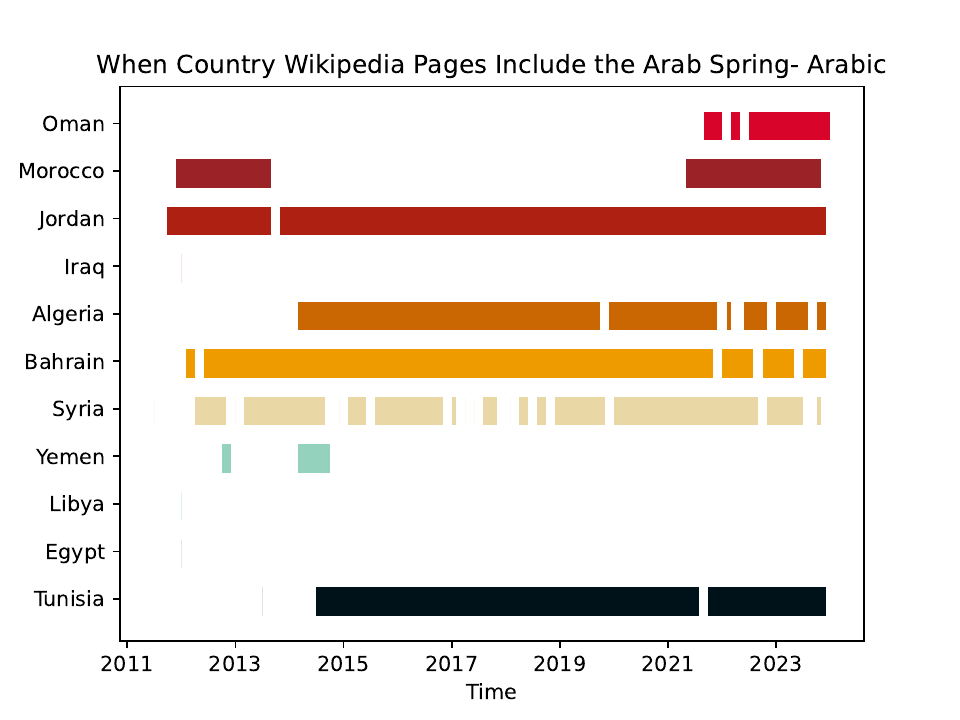}
        \caption{Country articles.}
    \end{subfigure} \hfill
    \begin{subfigure}{.49\textwidth}
        \includegraphics[width=\columnwidth]{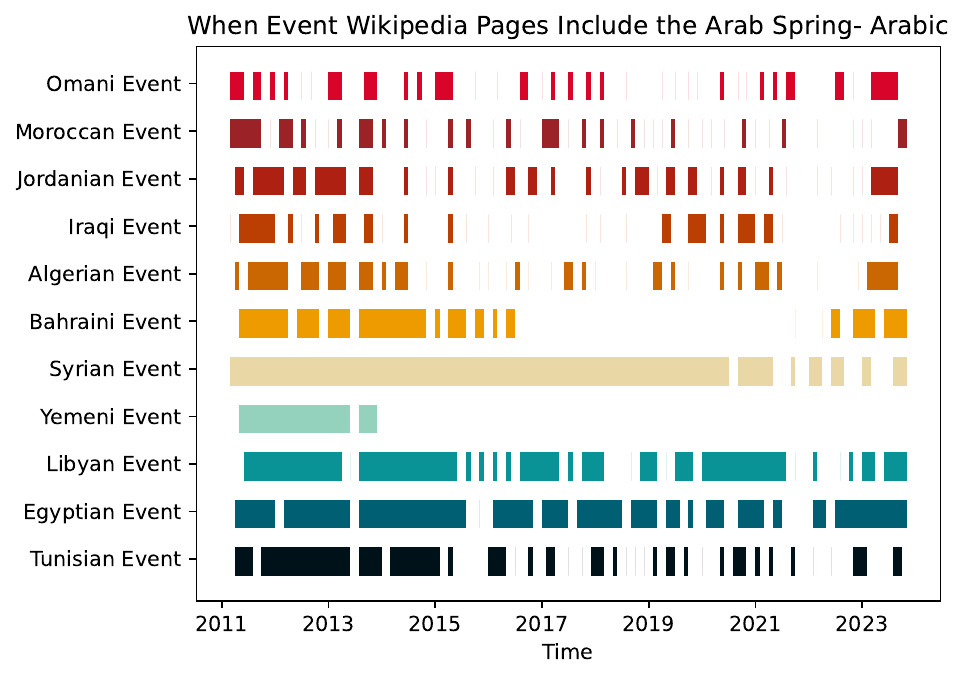}
        \caption{Event articles.}
    \end{subfigure}
    \caption{The first graph is a visual of the temporal inclusion of the 'Arab Spring' as an outlink in the Arabic language country articles. The second graph is the temporal inclusion of the 'Arab Spring' in the Arabic language event articles. This presents a perspective on the consolidation of the Arab Spring as a contextualization for the country and event Wikipedia articles.}
    \label{fig:temporal_inclusion_ar}
\end{figure*}

The Arabic language country and event articles debate the inclusion of the Arab Spring from the beginning, as shown in Figure 8.
The 'Libya' article and 'Iraq' article never include the 'Arab Spring' and the 'Egypt; article only includes it for 2 or 3 months at a time; so rarely that it did not show on our visual. Arabic editors did not find linking the 'Arab Spring' to be a necessary contextualization of these countries. 'Jordan' and 'Bahrain', articles about two countries that had relatively smaller events, consolidate the Arab Spring as context in their Wikipedia articles through the most time. 'Morocco' and 'Oman' have larger debates and then only as of 2021 include the outlink. 
The Syrian country article and event includes the Arab Spring through the longest amount of time but even this article in 2020 began debating its inclusion.
Opposing the English language community on Wikipedia, the Arabic language community found it hardly necessary to link the Arab Spring article in adjacent articles. 

There is a strong consolidation of the Arab Spring within the English country Wikipedia country articles but a majority of other articles analyzed across languages heavily debated the Arab Spring as necessary contextualization pointing to a varying consolidation within collective memory processes on Wikipedia and across languages. 

\section{Discussion}


How do collective memory processes unfold differently across different Wikipedia language editions? Previous research has looked at collective memory processes or compared content across language editions, but this paper provides novel empirical findings about both of these components. We proposed an ensemble of quantitative constructs for measuring four distinct collective memory processes identified in memory studies: salience, deliberation, contextualization, and consolidation. We evaluated these processes using a research design combining variation over time and across languages for articles related to the 2011 Arab Spring.

\textit{Salience} was operationalized as a measure of the size and number of outlinks. The English article surprisingly shows greater variation in size and content over time than the Arabic article, following patterns similar to punctuated equilibria: periods of stability interrupted by sudden changes. In contrast, the Arabic article is both shorter and more stable. \textit{Deliberation} was operationalized as similarity in the temporal vectors of outlink occurrence. The English article had five distinctive clusters of behavior relating to outlink inclusion that we classified as ``stable'', ``debated'', and ``forgotten'' interpretations while the Arabic article had only two clusters. \textit{Contextualization} was operationalized as a measure of overlap in the inter-language link graph. Despite the availability of relevant topics across languages, the Arabic and English articles referenced very different concepts. Finally, \textit{consolidation} was operationalized as references to the Arab Spring article in the articles about countries and their national-level protests. The Arabic Spring article is linked much less in Arabic articles for countries and events than in English.

These results highlighting disparities in the content and dynamics about the Arab Spring reinforce prior findings about the differences in coverage about the same topics across Wikipedia's language editions. The greater salience, deliberation, contextualization, and consolidation on the English compared to the Arabic article for these major events is a surprising counterpoint to the ``self-focus bias'' found in other multilingual analyses~\citep{hecht_measuring_2009,hecht_tower_2010}. The English article employs links reflecting Western cultural biases about history and politics that do not appear in the the Arabic version that provides its own distinctive contexts. Despite the greater geographic and cultural proximity of the events of the Arab Spring to Arabic speakers, the English-language article has far greater dynamism in its content and relationships than the Arabic-language article. The comparative absence of links to the \RL{الربيع العربي} (``Arab Spring'') article across Arabic-language articles about national histories and even national events (Figure~\ref{fig:temporal_inclusion_ar} is particularly surprising.

   


\subsection{Implications, limitations, and future work}
The Arabic and English Wikipedias' coverage of the Arab Spring are structured by distinct collective memory practices. We developed an ensemble of generalizable quantitative methods grounded in definitions from memory studies to enable comparisons of collective memory processes across languages and over time. These methods could be extended to other cases to examine whether these processes on the English Wikipedia are reliably distinctive because of its size and influence or if other language editions with similar cultural proximity to these events (Farsi, Hebrew, Turkish, \textit{etc}.) exhibit similarities to each other rather than to an outlier like English. Similarly, alternative constructs for collective memory processes and their empirical operationalizations might identify greater similarities or stronger differences.

These findings also contribute revisiting theories and methods for understanding collective memory processes. The concept of collective memory has been criticized from historiographical and cultural studies perspectives for reproducing dominant or revisionist historical narratives over alternative interpretations and marginalized perspectives~\cite{confino_CollectiveMemoryCultural_1997,kansteiner_FindingMeaningMemory_2002}. Wikipedia's rules governing citations emphasize the use of reliable sources which may similarly privilege some framings and interpretations over others. Although major contemporary historical events like the Arab Spring should attract more representative cross-sections of editorial perspectives, Wikipedia is not immune from motivated editors ``capturing'' pages to enforce specific narratives~\cite{grabowski_WikipediaIntentionalDistortion_2023,kharazian_GovernanceCaptureSelfGoverning_2023}. Future work should examine whether the ``afterlives'' of articles documenting major historical events are prone to capture or unreliable narratives because of declining editorial attention.

Our research design did not analyze the structure, dynamics, or content of discussions that accompanied the collaborations generating these articles. English and Arabic are each global languages with colonial legacies that do not map as cleanly to national boundaries as languages like Swedish, Hindi, or Japanese, for example. Collective memory cases drawn from those context may exhibit stronger self-focus biases than we saw with Arabic and English. Mechanisms like overlaps among contributors across articles and languages, conflicts over content, and processes for resolving disputes and forming consensus could have influenced the decisions to include or remove links over time we observed. A closer analysis of the content, discussions, and editors' activity could provide a richer description of additional forces shaping the content including and beyond outlinks on these articles.

The clustering of outlinks based on similarities in their temporal inclusion ignores other constructs for measuring the similarity of links and articles over time. Word embeddings trained on the content of out-linked articles, coauthorship patterns of contributing editors, or similarities in the pageview activity could all provide additional context and counterfactuals for the patterns of the observed temporal clustering. Understanding how outlink temporal inclusion vectors align with these other measures of similarity could have implications for large language and other natural language models trained on Wikipedia data. Because different language editions encode different content, content differences could be mistakenly equated and the resulting biases amplified. Characterizing the (mis)alignments across languages and over time is an important for identifying and preventing misalignments.
\bibliography{CMbibliography}

\begin{thebibliography}{27}
\providecommand{\natexlab}[1]{#1}

\bibitem[{Adar, Skinner, and Weld(2009)}]{adar_information_2009}
Adar, E.; Skinner, M.; and Weld, D.~S. 2009.
\newblock Information arbitrage across multi-lingual {Wikipedia}.
\newblock In \emph{Proceedings of the {Second} {ACM} {International} {Conference} on {Web} {Search} and {Data} {Mining}}, {WSDM} '09, 94--103. New York, NY, USA: Association for Computing Machinery.
\newblock ISBN 978-1-60558-390-7.

\bibitem[{Al-Shehari and Al-Sharafi(2022)}]{al-shehari_negotiating_2022}
Al-Shehari, K.; and Al-Sharafi, A.~G. 2022.
\newblock Negotiating {Wikipedia} narratives about the {Yemeni} crisis: {Who} are the alleged supporters of the {Houthis}?
\newblock \emph{Media, War \& Conflict}, 15(2): 183--201.
\newblock Publisher: SAGE Publications.

\bibitem[{Confino(1997)}]{confino_CollectiveMemoryCultural_1997}
Confino, A. 1997.
\newblock Collective Memory and Cultural History: Problems of Method.
\newblock \emph{American Historical Review}.

\bibitem[{Dandala, Mihalcea, and Bunescu(2012)}]{dandala_towards_2012}
Dandala, B.; Mihalcea, R.; and Bunescu, R. 2012.
\newblock Towards building a multilingual semantic network: identifying interlingual links in {Wikipedia}.
\newblock In \emph{Proceedings of the {First} {Joint} {Conference} on {Lexical} and {Computational} {Semantics} - {Volume} 1: {Proceedings} of the main conference and the shared task, and {Volume} 2: {Proceedings} of the {Sixth} {International} {Workshop} on {Semantic} {Evaluation}}, {SemEval} '12, 30--37. USA: Association for Computational Linguistics.

\bibitem[{Ferron and Massa(2011{\natexlab{a}})}]{ferron_collective_2011}
Ferron, M.; and Massa, P. 2011{\natexlab{a}}.
\newblock Collective memory building in {Wikipedia}: the case of {North} {African} uprisings.
\newblock In \emph{Proceedings of the 7th {International} {Symposium} on {Wikis} and {Open} {Collaboration}}, {WikiSym} '11, 114--123. New York, NY, USA: Association for Computing Machinery.
\newblock ISBN 978-1-4503-0909-7.

\bibitem[{Ferron and Massa(2011{\natexlab{b}})}]{ferron_wikirevolutions_2011}
Ferron, M.; and Massa, P. 2011{\natexlab{b}}.
\newblock {WikiRevolutions}: {Wikipedia} as a {Lens} for {Studying} the {Real}-time {Formation} of {Collective} {Memories} of {Revolutions}.
\newblock \emph{International Journal of Communication}, 1313--1332.

\bibitem[{Ferron and Massa(2014)}]{ferron_beyond_2014}
Ferron, M.; and Massa, P. 2014.
\newblock Beyond the Encyclopedia: Collective Memories in Wikipedia.
\newblock \emph{Memory Studies}, 7(1): 22--45.

\bibitem[{Ford(2022)}]{ford_writing_2022}
Ford, H. 2022.
\newblock \emph{Writing the {Revolution}: {Wikipedia} and the {Survival} of {Facts} in the {Digital} {Age}}.
\newblock MIT Press.

\bibitem[{Grabowski and Klein(2023)}]{grabowski_WikipediaIntentionalDistortion_2023}
Grabowski, J.; and Klein, S. 2023.
\newblock Wikipedia's Intentional Distortion of the History of the Holocaust.
\newblock \emph{The Journal of Holocaust Research}, 37(2): 133--190.

\bibitem[{Halbwachs(1992)}]{halbwachs_collective_1992}
Halbwachs, M. 1992.
\newblock \emph{On {Collective} {Memory}}.
\newblock Chicago: University of Chicago Press, 1st edition edition.
\newblock ISBN 978-0-226-11596-2.

\bibitem[{Hale(2014)}]{hale_multilinguals_2014}
Hale, S.~A. 2014.
\newblock Multilinguals and {Wikipedia} editing.
\newblock In \emph{Proceedings of the 2014 {ACM} conference on {Web} science}, {WebSci} '14, 99--108. New York, NY, USA: Association for Computing Machinery.
\newblock ISBN 978-1-4503-2622-3.

\bibitem[{He et~al.(2018)He, Lin, Adar, and Hecht}]{he_the_tower_of_babeljpg_2018}
He, S.; Lin, A.~Y.; Adar, E.; and Hecht, B. 2018.
\newblock The\_Tower\_of\_Babel.jpg: {Diversity} of {Visual} {Encyclopedic} {Knowledge} {Across} {Wikipedia} {Language} {Editions}.
\newblock \emph{Proceedings of the International AAAI Conference on Web and Social Media}, 12(1).
\newblock Number: 1.

\bibitem[{Hecht and Gergle(2009)}]{hecht_measuring_2009}
Hecht, B.; and Gergle, D. 2009.
\newblock Measuring self-focus bias in community-maintained knowledge repositories.
\newblock In \emph{Proceedings of the fourth international conference on {Communities} and technologies}, C\&amp;{T} '09, 11--20. New York, NY, USA: Association for Computing Machinery.
\newblock ISBN 978-1-60558-713-4.

\bibitem[{Hecht and Gergle(2010)}]{hecht_tower_2010}
Hecht, B.; and Gergle, D. 2010.
\newblock The tower of {Babel} meets web 2.0: user-generated content and its applications in a multilingual context.
\newblock In \emph{Proceedings of the {SIGCHI} {Conference} on {Human} {Factors} in {Computing} {Systems}}, {CHI} '10, 291--300. New York, NY, USA: Association for Computing Machinery.
\newblock ISBN 978-1-60558-929-9.

\bibitem[{Hickman et~al.(2021)Hickman, Pasad, Sanghavi, Thebault-Spieker, and Lee}]{hickman_understanding_2021}
Hickman, M.~G.; Pasad, V.; Sanghavi, H.~K.; Thebault-Spieker, J.; and Lee, S.~W. 2021.
\newblock Understanding {Wikipedia} {Practices} {Through} {Hindi}, {Urdu}, and {English} {Takes} on an {Evolving} {Regional} {Conflict}.
\newblock \emph{Proceedings of the ACM on Human-Computer Interaction}, 5(CSCW1): 34:1--34:31.

\bibitem[{Kansteiner(2002)}]{kansteiner_FindingMeaningMemory_2002}
Kansteiner, W. 2002.
\newblock Finding Meaning in Memory: A Methodological Critique of Collective Memory Studies.
\newblock \emph{History and Theory}, 41(2): 179--197.

\bibitem[{Keegan(2019)}]{keegan_dynamics_2019}
Keegan, B.~C. 2019.
\newblock The {Dynamics} of {Peer}-{Produced} {Political} {Information} {During} the 2016 {U}.{S}. {Presidential} {Campaign}.
\newblock \emph{Proceedings of the ACM on Human-Computer Interaction}, 3(CSCW): 33:1--33:20.

\bibitem[{Kharazian, Starbird, and Hill(2023)}]{kharazian_GovernanceCaptureSelfGoverning_2023}
Kharazian, Z.; Starbird, K.; and Hill, B.~M. 2023.
\newblock Governance Capture in a Self-Governing Community: A Qualitative Comparison of the Serbo-Croatian Wikipedias.
\newblock arxiv:2311.03616.

\bibitem[{Luyt(2016)}]{luyt_wikipedia_2016}
Luyt, B. 2016.
\newblock Wikipedia, collective memory, and the {Vietnam} war.
\newblock \emph{Journal of the Association for Information Science and Technology}, 67(8): 1956--1961.

\bibitem[{Massa and Scrinzi(2012)}]{massa_manypedia_2012}
Massa, P.; and Scrinzi, F. 2012.
\newblock Manypedia: comparing language points of view of {Wikipedia} communities.
\newblock In \emph{Proceedings of the {Eighth} {Annual} {International} {Symposium} on {Wikis} and {Open} {Collaboration}}, {WikiSym} '12, 1--9. New York, NY, USA: Association for Computing Machinery.
\newblock ISBN 978-1-4503-1605-7.

\bibitem[{Pentzold(2009)}]{pentzold_fixing_2009}
Pentzold, C. 2009.
\newblock Fixing the floating gap: {The} online encyclopaedia {Wikipedia} as a global memory place.
\newblock \emph{Memory Studies}, 2(2): 255--272.
\newblock Publisher: SAGE Publications.

\bibitem[{Porter, Krafft, and Keegan(2020)}]{porter_visual_2020}
Porter, E.; Krafft, P.~M.; and Keegan, B. 2020.
\newblock Visual {Narratives} and {Collective} {Memory} across {Peer}-{Produced} {Accounts} of {Contested} {Sociopolitical} {Events}.
\newblock \emph{ACM Transactions on Social Computing}, 3(1): 1--20.

\bibitem[{Roy, Bhatia, and Jain(2020)}]{roy_topic-aligned_2020}
Roy, D.; Bhatia, S.; and Jain, P. 2020.
\newblock A {Topic}-{Aligned} {Multilingual} {Corpus} of {Wikipedia} {Articles} for {Studying} {Information} {Asymmetry} in {Low} {Resource} {Languages}.
\newblock In Calzolari, N.; Béchet, F.; Blache, P.; Choukri, K.; Cieri, C.; Declerck, T.; Goggi, S.; Isahara, H.; Maegaard, B.; Mariani, J.; Mazo, H.; Moreno, A.; Odijk, J.; and Piperidis, S., eds., \emph{Proceedings of the {Twelfth} {Language} {Resources} and {Evaluation} {Conference}}, 2373--2380. Marseille, France: European Language Resources Association.
\newblock ISBN 979-10-95546-34-4.

\bibitem[{Roy, Bhatia, and Jain(2022)}]{roy_information_2022}
Roy, D.; Bhatia, S.; and Jain, P. 2022.
\newblock Information asymmetry in {Wikipedia} across different languages: {A} statistical analysis.
\newblock \emph{Journal of the Association for Information Science and Technology}, 73(3): 347--361.

\bibitem[{Twyman, Keegan, and Shaw(2017)}]{twyman_black_2017}
Twyman, M.; Keegan, B.~C.; and Shaw, A. 2017.
\newblock Black {Lives} {Matter} in {Wikipedia}: {Collective} {Memory} and {Collaboration} around {Online} {Social} {Movements}.
\newblock In \emph{Proceedings of the 2017 {ACM} {Conference} on {Computer} {Supported} {Cooperative} {Work} and {Social} {Computing}}, {CSCW} '17, 1400--1412. New York, NY, USA: Association for Computing Machinery.
\newblock ISBN 978-1-4503-4335-0.

\bibitem[{Yasseri, Gildersleve, and David(2022)}]{yasseri_chapter_2022}
Yasseri, T.; Gildersleve, P.; and David, L. 2022.
\newblock Chapter 9 - {Collective} memory in the digital age.
\newblock In O'Mara, S.~M., ed., \emph{Progress in {Brain} {Research}}, volume 274 of \emph{Collective {Memory}}, 203--226. Elsevier.

\bibitem[{Zubrzycki and Woźny(2020)}]{zubrzycki_comparative_2020}
Zubrzycki, G.; and Woźny, A. 2020.
\newblock The {Comparative} {Politics} of {Collective} {Memory}.
\newblock \emph{Annual Review of Sociology}, 46(1): 175--194.

\end{thebibliography}

\end{document}